\documentclass[conference,compsoc]{IEEEtran} 
\IEEEoverridecommandlockouts
\usepackage{cite}
\usepackage{amsmath,amssymb,amsfonts}
\usepackage{algorithm}
\usepackage{algorithmic}
\usepackage{graphicx}
\usepackage{textcomp}
\usepackage{xcolor}
\usepackage[utf8]{inputenc}
\usepackage{orcidlink}
\usepackage{stfloats}
\hypersetup{hidelinks}
\usepackage{subcaption}
\usepackage{caption}

\makeatletter
\def\@IEEEsectpunct{ }
\makeatother

\begin{document}
\title{\texttt{Onnx2TEE}: Automated Graph-Aware Model Partitioning for TEE-based Edge AI Protection

\thanks{Notice: This manuscript has been authored by UT-Battelle, LLC under Contract No. DE-AC05-00OR22725 with the U.S. Department of Energy and by Battelle Energy Alliance under U.S. Department of Energy (DOE) under Contract DE-AC07-05ID14517. The United States Government retains and the publisher, by accepting the article for publication, acknowledges that the United States Government retains a non-exclusive, paid-up, irrevocable, world-wide license to publish or reproduce the published form of this manuscript, or allow others to do so, for United States Government purposes. The Department of Energy will provide public access to these results of federally sponsored research in accordance with the DOE Public Access Plan (http://energy.gov/downloads/doe-public-access-plan).}
}

\author{
\IEEEauthorblockN{Daniela M. Vela Hernandez} 
\IEEEauthorblockA{\textit{Tennessee Technological University}\\
Cookeville, TN, USA \\
dmvelahern42@tntech.edu}
\and
\IEEEauthorblockN{Samuel Hildebrand} 
\IEEEauthorblockA{\textit{Oak Ridge National Laboratory}\\
Oak Ridge, TN, USA \\
hildebrandsc@ornl.gov}
\and
\IEEEauthorblockN{Sean Oesch} 
\IEEEauthorblockA{\textit{Oak Ridge National Laboratory}\\
Oak Ridge, TN, USA \\
oeschts@ornl.gov}
\and
\IEEEauthorblockN{Brian Weber} 
\IEEEauthorblockA{\textit{Oak Ridge National Laboratory} \\
Oak Ridge, TN, USA \\
weberb@ornl.gov}
\and
\IEEEauthorblockN{Shayan Monadjemi} 
\IEEEauthorblockA{\textit{Oak Ridge National Laboratory}\\
Oak Ridge, TN, USA \\
monadjemis@ornl.gov}
}
\maketitle

\begin{abstract}
AI models are increasingly being deployed at the edge. Since these models often require significant research and development to produce, are often deployed in potentially hostile environments, and downstream systems often rely on accurate data; protecting them from theft or manipulation is worthy of serious consideration. Trusted Execution Environments (TEEs), which provide hardware-enforced separation of software execution, can protect model weights but impose strict resource constraints, making full-model execution impractical for large neural networks. We present Onnx2TEE, an automated, model-independent pipeline that partitions ONNX models across rich, trusted execution environments for non-sequential feed-forward networks using graph-aware segmentation to preserve computational dependencies while minimizing boundary communication and reducing TEE-induced inference latency. Evaluation on an NVIDIA Jetson AGX Orin using an Arm TrustZone TEE demonstrates that model parameters can be highly concentrated in individual segments, supporting selective protection of parameter-dense model components. The evaluation further shows that the generated trusted app's memory requirements depend on both computation and intermediate tensor sizes.
\end{abstract}


\section{Introduction}\label{sec:introduction}
As more sophisticated computing moves to edge devices, particularly AI, the substantial engineering and infrastructure costs required to develop models incentivize attacks that target the exfiltration of valuable model weights directly from the execution environment. Additionally, many downstream systems rely on the presumed accuracy of an edge model. An attacker may be incentivized to manipulate the weights of an edge model to circumvent restrictions or gain an advantage in a downstream system. 

Since users on an edge device can view local model weights decrypted at inference time in the process's virtual address space by use of compromised privileged software such as the operating system (OS), protecting model weights during inference therefore requires preventing untrusted software on the device from accessing the portions of model parameters used during execution.
There is significant interest in running model inference utilizing Trusted Execution Environments (TEEs) that provide hardware-enforced separation between normal, rich software execution (REE) and a secure, authenticated execution (TEE).
Due to the resource-constrained nature of TEEs, running modern models entirely in trusted execution is often infeasible.
Thus, several approaches have evolved to utilize split inference across execution environments to address this constraint by partitioning models to ideally execute security-sensitive portions using the TEE while leaving less sensitive computation in the REE. \cite{mo_darknetz_2020, gangal_hybridtee_2020, hu_efftee_2025, steiakakis_trusted_2026}

Because feed-forward deep neural networks can contain branching and merging dependencies, partitioning a model between the REE and TEE arbitrarily selecting a single layer boundary and passing only the preceding layer's output is inefficient. \cite{EdgeFlow, qararyah2021computational, tarnawski2020efficient} 
While related works employ strategies such as pruning models  \cite{hu_efftee_2025} to fit into the TEE, offloading computation to a remote TEE \cite{gangal_hybridtee_2020}, or increasing TEE capacity to accommodate heavyweight runtimes \cite{steiakakis_trusted_2026}, these approaches either modify the model, introduce additional infrastructure, or expand the trusted computing base.
A model-independent, computational-graph-aware partitioning framework that adapts model execution to existing TEE resource constraints remains unexplored. 

We introduce \textbf{\texttt{Onnx2TEE}}, an automated deployment pipeline that transforms an ONNX-format \cite{onnx} model into a TEE-enabled split-inference application for modern, non-sequential feed-forward networks. 
To do this, we define a segmentation algorithm that preserves dependencies in a model's computational graph and minimizes REE/TEE data communications, yielding candidate partition points for execution environment switching. Furthermore,
\textbf{\texttt{Onnx2TEE}} assesses candidate partitions for memory requirements and automatically generates software that utilizes the TEE for inference, easing developer effort in model TEE integration.

The code for \textbf{\texttt{Onnx2TEE}} will be released upon acceptance, pending approval from the funding organization.

 Testing the \textbf{\texttt{Onnx2TEE}} framework, we compare inference of a small CNN-based MNIST image classifier and \texttt{yolo26nano}, representative of an architecturally sophisticated model, to answer the following research questions. We evaluate \textbf{\texttt{Onnx2TEE}} using the Arm TrustZone TEE on an NVIDIA Jetson AGX Orin Developer Kit.
\begin{itemize}
    \item \textbf{RQ1:} How does Onnx2TEE's model segmentation algorithm affect the distribution of trained weights for the
aforementioned models?
    \item \textbf{RQ2: } How do different REE and TEE environment partition configurations affect the heap requirement of the generated trusted application?
    \item \textbf{RQ3:} How do different partition configurations affect the
inference time of a model?
\end{itemize}

The evaluation of our graph-aware method shows interesting findings in the trade-offs between model-weight protection, TEE memory, and inference latency: model parameters may concentrate in individual segments, TEE memory depends on computation and intermediate tensor sizes, and TEE execution introduces measurable inference overhead.

\section{Background}\label{sec:danny_background}

\subsection{Trusted Execution Environments}
Trusted Execution Environments (TEEs) are hardware-enforced security solutions that isolate sensitive data and applications from software-based attacks, including those originating from privileged operating systems in the Rich Execution Environment (REE). They preserve the confidentiality and integrity of authenticated code for applications such as digital rights management, runtime integrity, and privacy-preserving data analysis \cite{globalplatform2018tee}.  
TEE-based systems commonly rely on mechanisms such as secure boot, which verifies TEE images before execution; attestation; isolation of TEE processes from the REE; and sealing secrets to local storage using keys derived from a trust anchor \cite{li_survey_2023}.
TEEs differ in architecture based on their deployment target (local, cloud, or edge environments), the hardware components that serve as trust anchors, isolation model (multi-TEE versus single-TEE), and support for peripheral protection \cite{jauernig_trusted_2020}.
Popular implementations include Intel's SGX and TDX, AMD's SEV, ARM's TrustZone, and academic systems such as Sanctum and Sanctuary \cite{intel_sgx_programming_reference_2014, SGXexplained, amd_sev, arm_trustzone_2009, costan2016sanctum}. TEEs that follow the GlobalPlatform standards for REE to TEE communication enable applications to be ported across vendors after re-signing with the target platform's root of trust \cite{globalplatform2018tee}.
Attacks on TEEs are commonly categorized as software, architectural, microarchitectural, or side-channel attacks, targeting TEE software, hardware, processor components like caches, or observable behaviors such as timing, network traffic, and memory access patterns \cite{munoz_survey_2023}.

\subsubsection{Arm Trustzone}
Rather than instantiating multiple security enclaves, ARM TrustZone is a single Trusted Execution Environment architecture for edge-device CPUs that provides two execution domains: Secure World and Non-Secure World. 
Trusted applications (TAs) execute within the secure world on top of a Trusted Operating System, while normal client applications (CAs) execute in the non-secure world.
A world switch is initiated when the operating system in the non-secure world invokes a secure monitor call, trapping execution to Arm Trusted Firmware Secure Monitor at the highest exception (EL3). The Arm Trusted Firmware saves CPU context, updates the SCR\_EL3 register's Non-Secure bit, and transfers control to the target environment.
Memory isolation is enforced through hardware mechanisms, such as the memory management unit, preventing non-secure world software from accessing secure world memory regions \cite{munoz_survey_2023}. 
However, TrustZone remains vulnerable to attacks targeting its trusted software stack, where vulnerabilities in the trusted operating system, trusted application, or Secure Monitor interfaces can allow a compromised non-secure world to affect the secure world \cite{cerdeira_sok_2020}. 

\subsection{Feedforward Deep Neural Networks}
AI model inference can be represented by a computational graph of form \begin{math}G=(V,E)\end{math}, where vertices represent computational operators and edges represent intermediate tensors exchanged between operators. \cite{abadi2016tensorflow} Operators may be non-trainable, performing deterministic operations such as activation functions, or trainable, additionally consuming learned parameters such as weights and biases.
Feed-forward deep neural networks (FDNNs) represent inference as a directed acyclic graph, allowing branching paths, merging operations, and skip connections. \cite{szegedy2015going,he2016deep} Unlike sequential models, where each layer has a single predecessor and successor, FDNNs may contain multi-input operators consuming multiple intermediate tensors. These non-sequential dependencies complicate layer-wise model partitioning, motivating graph-aware secure inference approaches.

\subsection{Model Weight Exfiltration on Edge Devices}
AI model weights represent valuable intellectual property due to the cost of data collection, training, and optimization. Model weight exfiltration occurs when an unauthorized entity obtains a provider's exact or copied model parameters from a storage or execution environment. Unlike model extraction, which reconstructs an approximate model through black-box queries \cite{tramer_stealing_nodate} or side-channel observations, weight exfiltration directly targets the deployed model parameters.
As inference moves from remote servers to edge devices for reduced latency and improved privacy, deployed models shift from a query-only black-box setting to an environment where compromised privileged software, such as the operating system, may access decrypted model weights during execution. Consequently, protections for models at rest are insufficient against runtime exfiltration attacks.

\subsection{TEE-Aware Model Partition and Execution Strategies }
DarkneTZ \cite{mo_darknetz_2020} introduced TEE-assisted DNN partitioning by executing
sensitive model components within the TEE, leaving remaining
computation in the Rich Execution Enviornment (REE). However, its approach assumes only the
final layer requires protection and performs layer-wise partitioning
using Darknet-based models, limiting applicability to modern DNN
architectures and ML frameworks.
HybridTEE \cite{gangal_hybridtee_2020} extended TEE partitioning by selecting privacy-aware
partition points and dividing inference into LocalNet, RemoteNet, and
PredNet components. While providing more flexible partitioning than
DarkneTZ, it is designed for vision workloads, offloads intermediate
computation to a remote TEE, and does not explicitly model
dependencies introduced by non-sequential computational graphs.
EffTEE \cite{hu_efftee_2025} enables fully local TEE inference through
structured pruning and memory-aware layer slicing. Its partitioning
strategy maximizes the number of layers executed within the TEE under
memory constraints. However, although evaluated on YOLO models,
EffTEE does not explicitly describe how layer-wise slicing preserves tensor
dependencies in non-sequential architectures. Additionally, its reliance on
Darknet-based models and DarkneTZ-generated TEE applications limits
framework portability.
An Intel SGX-based inference system \cite{steiakakis_trusted_2026} improved portability by
executing ONNX models within an enclave using a general-purpose ML
runtime. However, the resulting software stack introduces a large
trusted computing base that limits suitability for resource-constrained
edge TEEs.
Existing approaches therefore provide TEE execution through
partitioning, optimization, or runtime abstraction, but lack a
framework-independent, graph-aware  partitioning method for feed-forward DNNs while generating lightweight edge-deployable TEE applications.

\section{Onnx2TEE Implementation}\label{design}

\textbf{\texttt{Onnx2TEE}} addresses three challenges in TEE-based inference: framework independent model support, graph-aware partitioning with minimized Rich Execution Environment (REE) / Trusted Execution Environment (TEE) communication, and automated generation of GlobalPlatform-compliant non-secure world client applications (CA) and secure world trusted applications (TA). 
It transforms ONNX models into deployable TEE-enabled inference applications through an automated compilation pipeline.
Given an ONNX model, \textbf{\texttt{Onnx2TEE}} performs dependency-aware graph partitioning, generates model-specific inference code, integrates the generated code into reusable CA and TA templates, and validates deployment feasibility under TEE memory constraints. 
This removes the need for developers to manually implement model-specific trusted applications while enabling lightweight edge-device deployment.

\subsection{ONNX-Based Model Ingestion}

Open Neural Network Exchange (ONNX) is a framework-agnostic model interchange format that represents inference as computational graphs, where nodes correspond to operators applied to input tensors.
Since a single neural network layer may map to multiple ONNX nodes, \textbf{\texttt{Onnx2TEE}} uses a graph rather than layer abstraction for identifying dependencies.
Unlike DarkneTZ and EffTEE, \textbf{\texttt{Onnx2TEE}} adopts ONNX to support models exported from modern ML frameworks including PyTorch, TensorFlow, and Keras through a standardized operator set. 
Since ONNX defines computation but not execution, \textbf{\texttt{Onnx2TEE}} can parse the graph and generate model-specific inference code without embedding general-purpose inference engines like ONNX Runtime or TVM.

\subsection{Graph-Dependency-Aware Model Partitioning}\label{sec:seg_alg}
Existing TEE partitioning approaches often assume sequential layer execution, whereas modern DNNs are represented as feed-forward computational graphs containing skip connections, concatenations, and branching paths. 
Naive layer-wise partitioning can require multiple intermediate tensors to cross the TEE boundary, increasing communication overhead during secure monitor calls (SMCs). \textbf{\texttt{Onnx2TEE}} instead performs deterministic dependency-aware graph segmentation followed by environment-aware partitioning, preserving execution correctness while selecting partition boundaries that require the transfer of a single intermediate tensor between execution environments. 

Because subsequent inference code generation relies on static memory allocation, \textbf{\texttt{Onnx2TEE}} requires tensor dimensions to be resolved during offline compilation. The use of ONNX enables us to parse the model computational graph to verify it as an feed forward deep neural network (FDNN) with no unsupported operators such as recurrent structures.
We define a model segment as a single-entry, single-exit connected induced subgraph. Because ONNX edges represent intermediate tensors, each segment exchanges exactly one input tensor and one output tensor with adjacent segments. Model segmentation then decomposes a feed-forward computational graph into an ordered sequence of non-overlapping model segments. Beginning at the model input, segments are constructed greedily by repeatedly selecting the maximal model segment beginning at the current entry node. The entry node of each subsequent segment is the successor of the previous segment's exit node.
Let $S=(S_1,S_2,\ldots,S_n)$ denote the resulting ordered segmentation, where $S_i \cap S_j=\emptyset$ for $i\neq j$ and each $S_i$ is a model segment.
A model partition consists of one or more consecutive model segments $S_i,\ldots,S_j$ from the ordered segmentation $S$, where $i$ and $j$ denote the first and last segment indices, respectively:
\[
P = \bigcup_{k=i}^{j} S_k
\quad \text{where} \quad
1 \le i \le j \le n.
\]

Candidate partition points, $C=\{c_0,c_1,\ldots,c_{n}\}$ , include the model input and output as well as internal boundaries between consecutive segments. The internal partition point $c_i$ lies between consecutive segments $S_i$ and $S_{i+1}$ for $1\leq i<n$, while $c_0$ and $c_n$ correspond to the model input and output, respectively. Thus, a model with $n$ segments has $n+1$ possible partition points.

Given $G=(V,E)$, its ordered segmentation $S$, and the candidate partition points $C$, the partitioning algorithm presents valid partition points that preserve graph execution order and avoid multi-input dependencies, producing the ordered partitions $P_{preREE},P_{TEE},P_{postREE}$.
The resulting partitions are exported as independent ONNX submodels and compiled into the REE's client application and TEE's trusted application inference binaries.

\subsection{Native Code Generation}
Rather than embedding a general-purpose ONNX runtime, as in \cite{steiakakis_trusted_2026}, \textbf{\texttt{Onnx2TEE}} uses the emmtrix \cite{emmtrixemx-onnx-cgen_2026} ONNX-to-C ahead-of-time compiler to translate an ONNX graph into standalone C. Because the output is specialized to a single model, it includes only the required operators and execution logic rather than the generic infrastructure needed to support arbitrary models, substantially reducing the trusted code base. Model weights are embedded as compile-time C arrays, eliminating the need for an ONNX parser or runtime model loading.

\subsection{Generic TEE Execution Templates}
To avoid requiring developers to implement a custom TEE application for each model, \textbf{\texttt{Onnx2TEE}} provides reusable CA and TA templates that encapsulate common secure execution logic. 
The generated model inference code is integrated into the templates, while generic tensor interfaces allow the same execution framework to support different models. 
The templates enforce security-critical practices that are necessary for secure TEE application development. For example, to mitigate Time-of-Check to Time-of-Use (TOCTOU) attacks on shared memory between the secure and non-secure worlds, the TA template copies input tensors into private secure buffers before inference. By moving security-sensitive functionality into reusable templates, \textbf{\texttt{Onnx2TEE}} reduces both developer effort and the likelihood of model-specific TEE programming errors.

\subsection{Memory-Aware Deployment Validation}
Compilation success does not guarantee deployability on resource-constrained TEEs. \textbf{\texttt{Onnx2TEE}} validates candidate partitions by estimating the secure-world memory requirements of $P_{TEE}$ including embedded model weights, intermediate activations, and stack usage; and rejects configurations exceeding available TEE memory. Since model parameters are embedded into the resulting CA/TA artifacts, the execution partitioning strategy must be determined before compilation.

\subsection{Runtime Inference Execution}
After compilation, the user deploys the generated CA and TA binaries to the target edge device. During inference, the CA invokes the TA through the TEE interface to execute TEE-resident sub-models, transferring input tensors to the secure world and receiving inference results in return. For partitioned models, execution follows the compile-time selected REE/TEE ordering, with intermediate tensors exchanged between REE and TEE components as needed.

\section{Evaluation}
All edge device experiments were performed on NVIDIA's Jetson AGX Orin housing an ARM Cortex-A78AE CPU.
The Jetson's SDK was Jetpack v5.1.2 running Linux for Tegra r35.4.1. 
The CPU trusted execution environment utilizes ARM TrustZone hardware architecture with OPTEE v3.19 as the trusted operating system.
A x86\_64 Ubuntu 22.04.05 LTS virtual machine used the jetson-linux toolchain to cross-compile NVIDIA's provided OPTEE source packages to create client applicaiton and trusted application executables. 
The dedicated TEE RAM was kept to the default of 64 MiB.

\subsection{Model Selection}
\label{models}
To evaluate our pipeline, we use two computer vision models of differing complexity. 
As a simple image classifier, we trained a lightweight PyTorch Convolutional Neural Network (CNN) on the MNIST dataset, comprising two 3x3 convolutional layers (16 and 32 channels) with ReLU activations and max-pooling, followed by a 128-unit fully connected layer and a 10-class output layer. 
The model was trained for two epochs using Adam and cross-entropy loss, resulting in a network with 206,922 parameters. 
As a more complex model, we selected the nano variant of Ultralytic's YOLO26 architecture \cite{jocher2026ultralyticsyolo26unifiedrealtime}, a non-sequential object detection model with skip connections and branching, containing 2.4 million parameters and is designed for efficient edge deployment with end-to-end NMS-free inference. 
Both models were exported to ONNX using \verb|torch.onnx.export| with static input shapes and ONNX opset version 18.

\subsection{Methodology and Metrics}
The MNIST Classification CNN fits entirely within the TEE, while YOLO26n requires partitioned execution.
The CNN was evaluated on eight MNIST images, while YOLO26n was evaluated on the COCO8 dataset.
For each model, we evaluated all deployable Rich Execution Environment/Trusted Execution Environment partitions generated by \textbf{\texttt{Onnx2TEE}}, excluding configurations that exceeded the available TEE memory budget.
We measured (i) TEE memory footprint as Trusted Application heap usage, (ii) cross-world communication as shared-memory buffer size, and (iii) end-to-end inference latency using \texttt{clock\_gettime(CLOCK\_MONOTONIC)} in the client application immediately before and after inference. 
Reported latency includes secure monitor calls and world-switch overhead.
Shared-memory buffers are rounded up to the nearest secure page size, which on the evaluation platform is 2 MiB.
Both client application and trusted application were compiled using the \texttt{gcc -Os} optimization level to avoid performance differences attributed to compiler optimization. 

\subsection{Results}\label{results_section}
We examine three aspects of TEE-based model partitioning: the distribution of model parameters across generated segments, the TEE memory requirements of candidate partitions, and the inference latency associated with TEE execution. 
These metrics highlight the tradeoffs between model-weight protection, resource constraints, and inference performance when deploying models in edge device TEEs.

\subsubsection{RQ1: How does the segmentation algorithm described in Section \ref{sec:seg_alg} affect the distribution of trained weights for the aforementioned models?}

\begin{figure}
    \centering
    \includegraphics[width=0.5\linewidth]{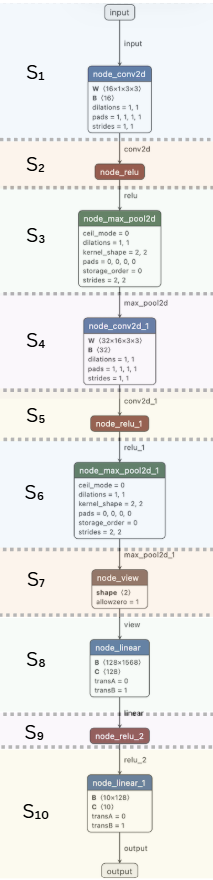}
    \caption{\textbf{Segmentation of MNIST Image Classifier Computational Graph to Preserve Dependencies \& Minimize REE/TEE Communications.}
    \normalfont Because the model is sequential, having no branches or skip connections, every node depends only on the output of the preceding node. The algorithm in \ref{sec:seg_alg} therefore splits the model after each node, resulting in segments $S_1, \cdots, S_{10}$ that each have the same number of nodes. Each colored region in the figure represents one model segment and its corresponding operations. Each segment can be independently assigned to the REE or TEE while preserving the model's dependencies.
    }
    \label{fig:mnist_seg}
\end{figure}



\begin{figure}
    \centering
    \begin{subfigure}{\linewidth}
        \centering
        \includegraphics[width=\linewidth]{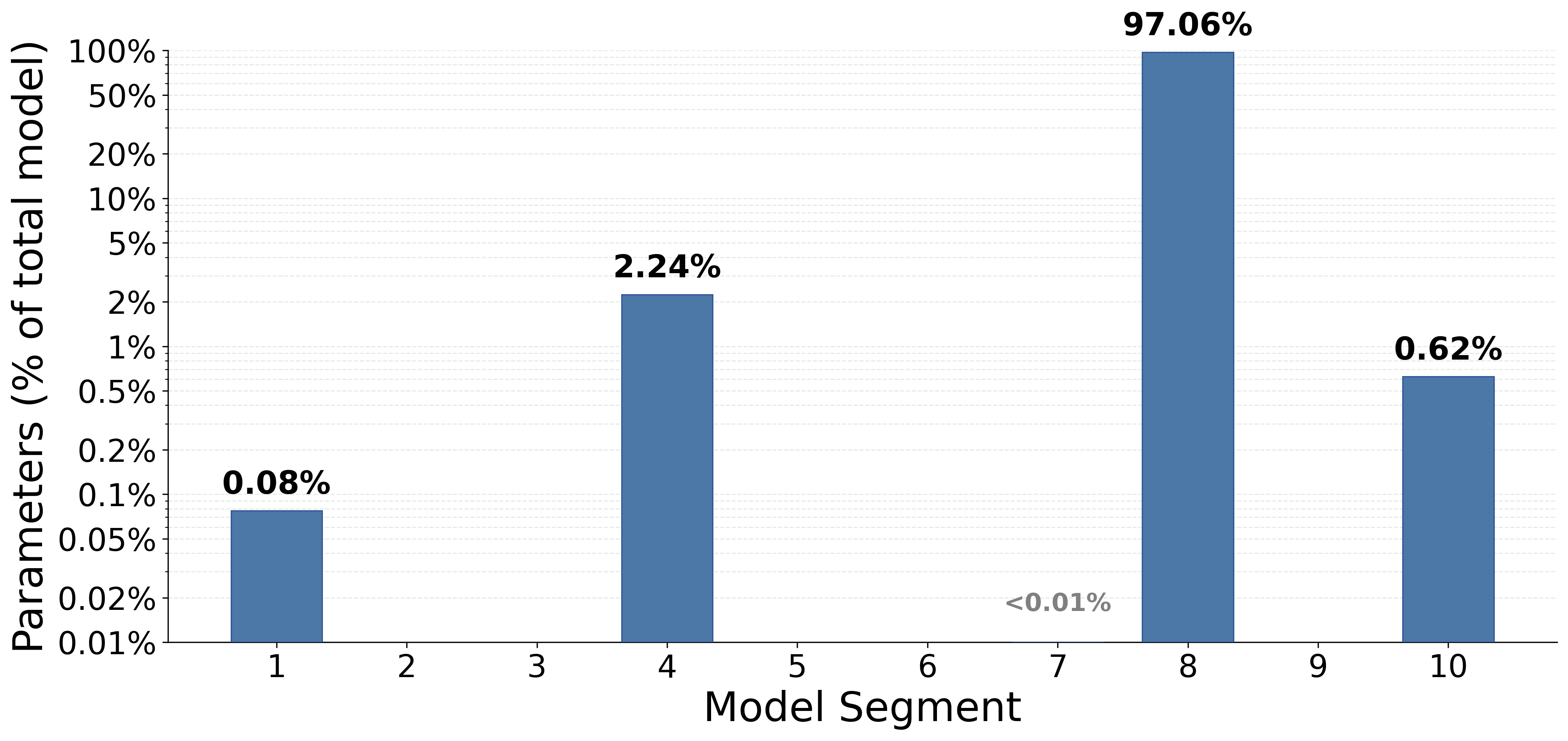}
        \caption{MNIST}
        \label{fig:mnist_seg_param_distr}
    \end{subfigure}

    \vspace{0.5em}
    \begin{subfigure}{\linewidth}
        \centering
        \includegraphics[width=\linewidth]{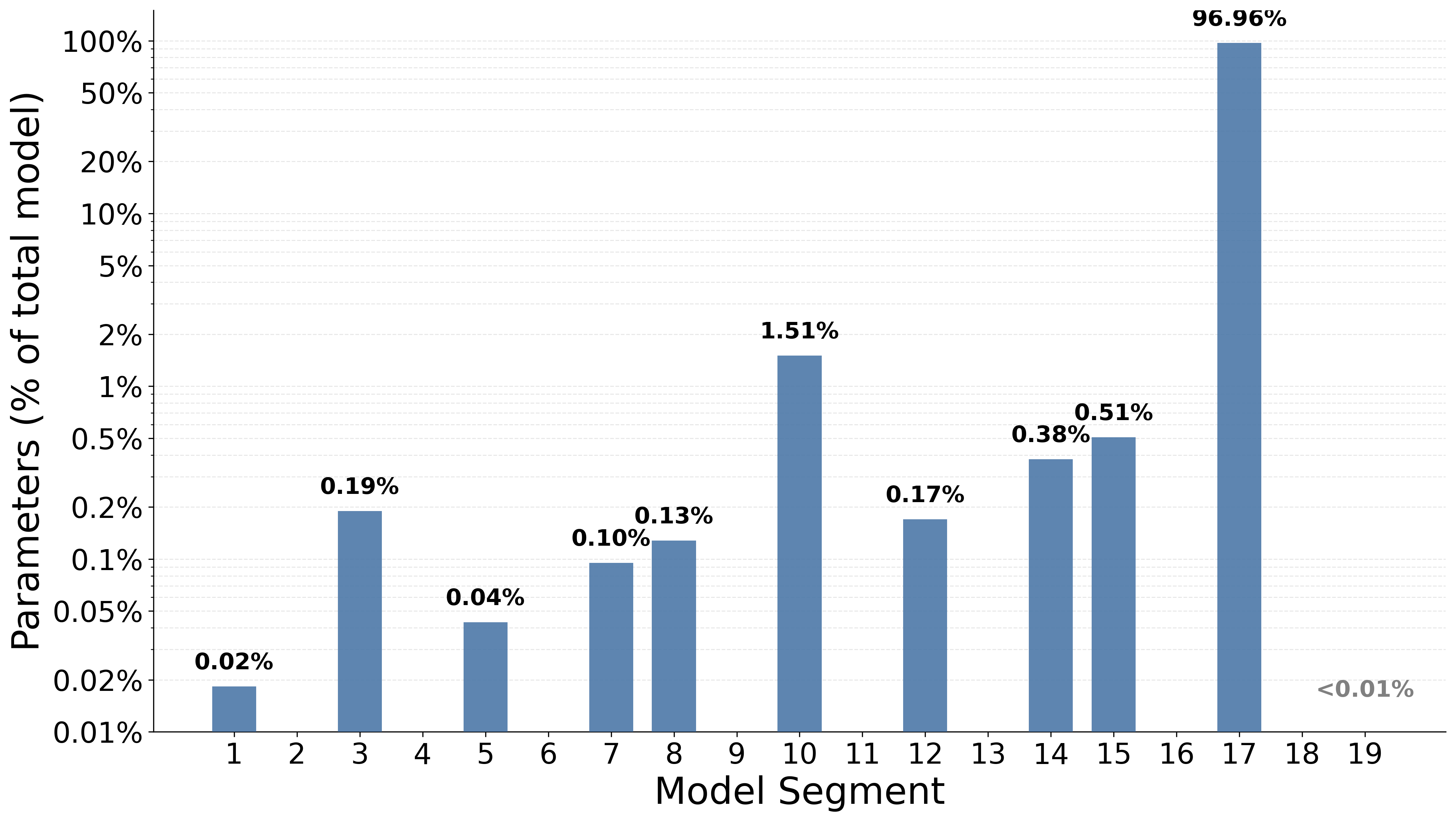}
        \caption{YOLO26n}
        \label{fig:yolo26n_seg_param_distr}
    \end{subfigure}

    \caption{\textbf{Distribution of Learned Parameters Across Model Segments for
    (a) MNIST and (b) YOLO26n.}
    \normalfont The y-axis uses a logarithmic scale to show parameter counts spanning multiple orders of magnitude. MNIST contains learned parameters in Segments
    $\{S_1,S_4,S_7,S_8,S_{10}\}$ of 10 segments, with 97.06\% concentrated in
    Segment $S_8$. YOLO26n contains learned parameters in Segments
    $\{S_1,S_3,S_5,S_7,S_8,S_{10},S_{12},S_{14},S_{15},S_{17},S_{19}\}$ of 19 segments, with 96.96\% of the learned parameters
    concentrated in Segment $S_{17}$.
    The differing distributions illustrate that parameter concentration is model-specific, making segment placement important for protecting model weights.
    }
    \label{fig:model_seg_param_distr}
\end{figure}

As shown in Figures ~\ref{fig:mnist_seg}, the segmentation algorithm described in Section \ref{sec:seg_alg} splits models according to their data-flow, creating a new segment where the architecture reduces to a single tensor to minimize Rich Execution Environment and Trusted Execution Environment communication. Consequently, segments are not necessarily balanced in the number of operators (nodes) or parameters (trainable nodes) they contain.
The sequential MNIST classifier can be segmented after individual nodes, resulting in ten segments. In contrast, skip connections and branches throughout YOLO26n constrain its splitting and result in nineteen segments. Figures ~\ref{fig:mnist_seg_param_distr} and ~\ref{fig:yolo26n_seg_param_distr} show that this architectural difference produces highly non-uniform parameter distributions. 
For MNIST, 97.06\% of the model's parameters are concentrated in Segment 8, primarily due to its fully connected layer. Similarly, 96.96\% of YOLO26n's parameters are amassed in Segment 17, which contains the neck and final feature-fusion operations preceding detection. Thus, increasing the number of model segments does not necessarily produce a balanced distribution of model parameters.
This has implications for mitigating model-weight exfiltration. Rather than prioritizing the number of segments placed in the TEE, parameter-heavy segments may provide greater protection. However, this must be assessed against the memory requirements of placing those segments and their associated weights within the TEE. 
This motivates the analysis of memory consumption across model segments in the subsequent question.

\subsubsection{RQ2: How do different partition configurations affect the TEE heap requirement of the generated trusted application?}


\begin{figure}
    \centering
    \label{fig:heap_heatmaps}
    
    \begin{subfigure}{\linewidth}
        \centering
        \includegraphics[width=\linewidth]{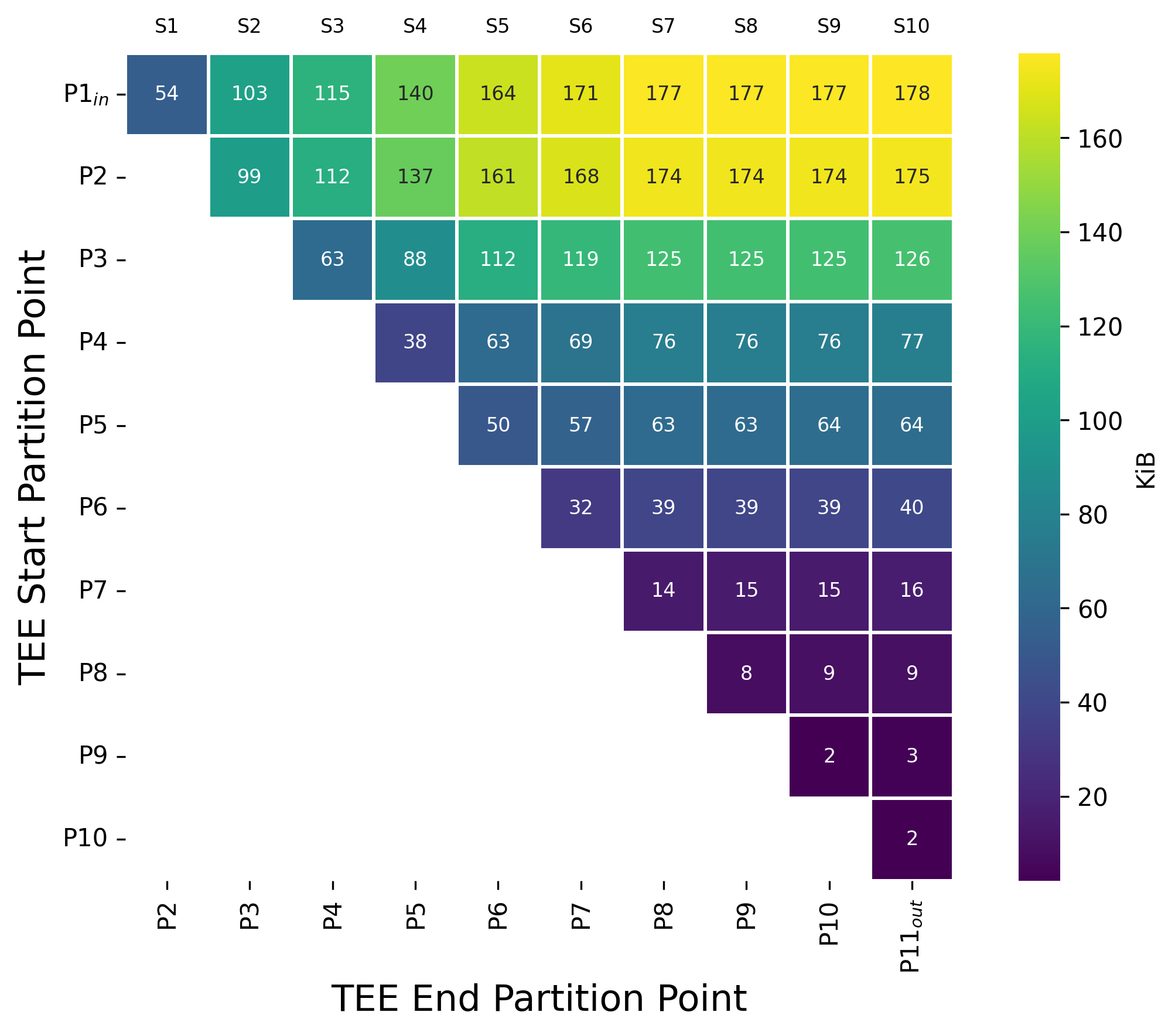}
        \caption{MNIST}
        \label{fig:MNIST_heap_heatmap}
    \end{subfigure}

    \vspace{0.5em}
    
    \begin{subfigure}{\linewidth}
        \centering
        \includegraphics[width=\linewidth]{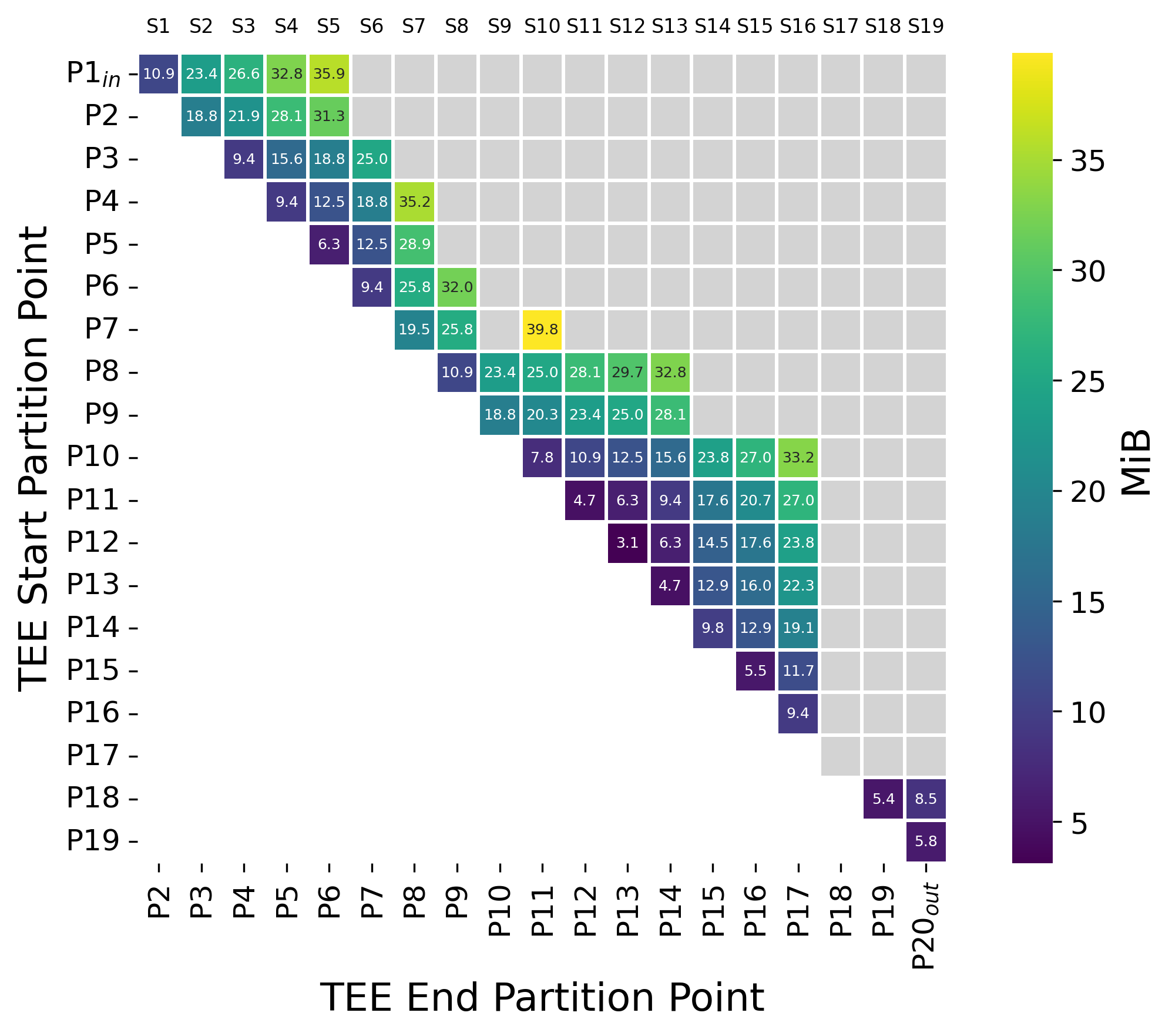}
        \caption{YOLO26n}
        \label{fig:YOLO26n_heap_heatmap}
    \end{subfigure}

    \caption{\textbf{Trusted Application Average Heap Usage Across TEE Partition Configurations for (a) MNIST and (b) YOLO26n.}
    \normalfont Rows and columns denote the TEE start and end partition points, respectively. 
    Consecutive partition points delimit model segments $S_1, \cdots, S_n$, such that a configuration $(P_i,P_j)$ places the model segments between $P_i$ and $P_j$ in the TEE. 
    Cell values report the trusted application heap requirement. 
    For MNIST, $P_1$ and $P_{11}$ correspond to the model input and output, respectively, and values are reported in KiB. 
    For YOLO26n, $P_1$ and $P_{20}$ correspond to the model input and output, respectively, values are reported in MiB, and gray cells in indicate valid partition configurations that were not deployable under the available TEE memory budget.
    TEE heap usage increases as the TEE encompasses more of the model.  YOLO26n exhibits a clear region of memory-infeasible configurations.
    }
\end{figure}

Figures ~\ref{fig:MNIST_heap_heatmap} and ~\ref{fig:YOLO26n_heap_heatmap} show \textbf{\texttt{Onnx2TEE}}'s trusted application heap requirement calculation from the deployment-validation procedure to quantify the TEE memory footprint of each partition configuration. \textbf{\texttt{Onnx2TEE}} accounts for model weights, temporary activation buffers, and the duplicated input/output buffers used to protect shared memory. Configurations whose estimated heap usage exceeds the available TEE memory are marked as undeployable.
For MNIST, all valid partition configurations fit within the available TEE memory, with heap usage generally increasing as more segments are placed in the TEE. YOLO26n exhibits substantially greater variation: several configurations exceed the memory budget despite containing fewer segments than some deployable configurations. This occurs because segments are not uniform in their memory-heavy parameter distribution, as shown in RQ1, and because boundary tensors contribute directly to the TEE memory footprint. 
For example, $(P_7,P_{11})$ is deployable whereas $(P_7,P_{10})$ is not, even though the former contains one additional segment. Extending the partition to $P{11}$ moved the TEE output boundary past a spatial downsampling operation, reducing the output tensor from $[1,64,160,160]$ to $[1,64,80,80]$. Consequently, the duplicated output buffer decreases from 6,400~KiB to 1,600~KiB, more than offsetting the additional intermediate activation memory.
Thus, TEE heap usage depends not simply on the number of segments placed in the TEE, but on the memory characteristics of the selected segments and the sizes of their boundary tensors.

\subsubsection{RQ3: How do different partition configurations affect the inference time of a model?}

\begin{figure}
    \centering

    \begin{subfigure}{\linewidth}
        \centering
        \includegraphics[width=\linewidth]{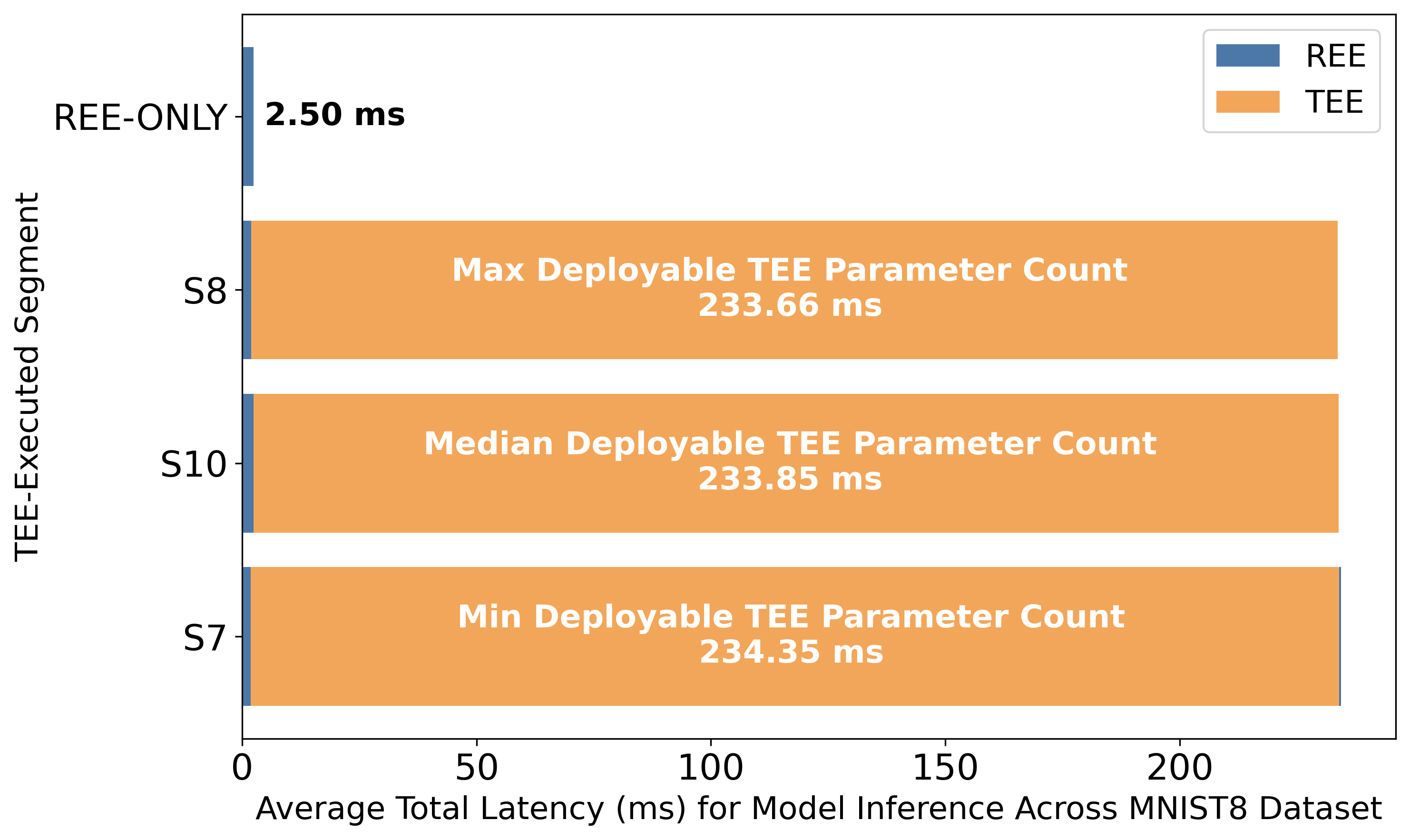}
        \caption{MNIST}
        \label{fig:MNIST_latency_representative_param_segs}
    \end{subfigure}

    \vspace{0.5em}

    \begin{subfigure}{\linewidth}
        \centering
        \includegraphics[width=\linewidth]{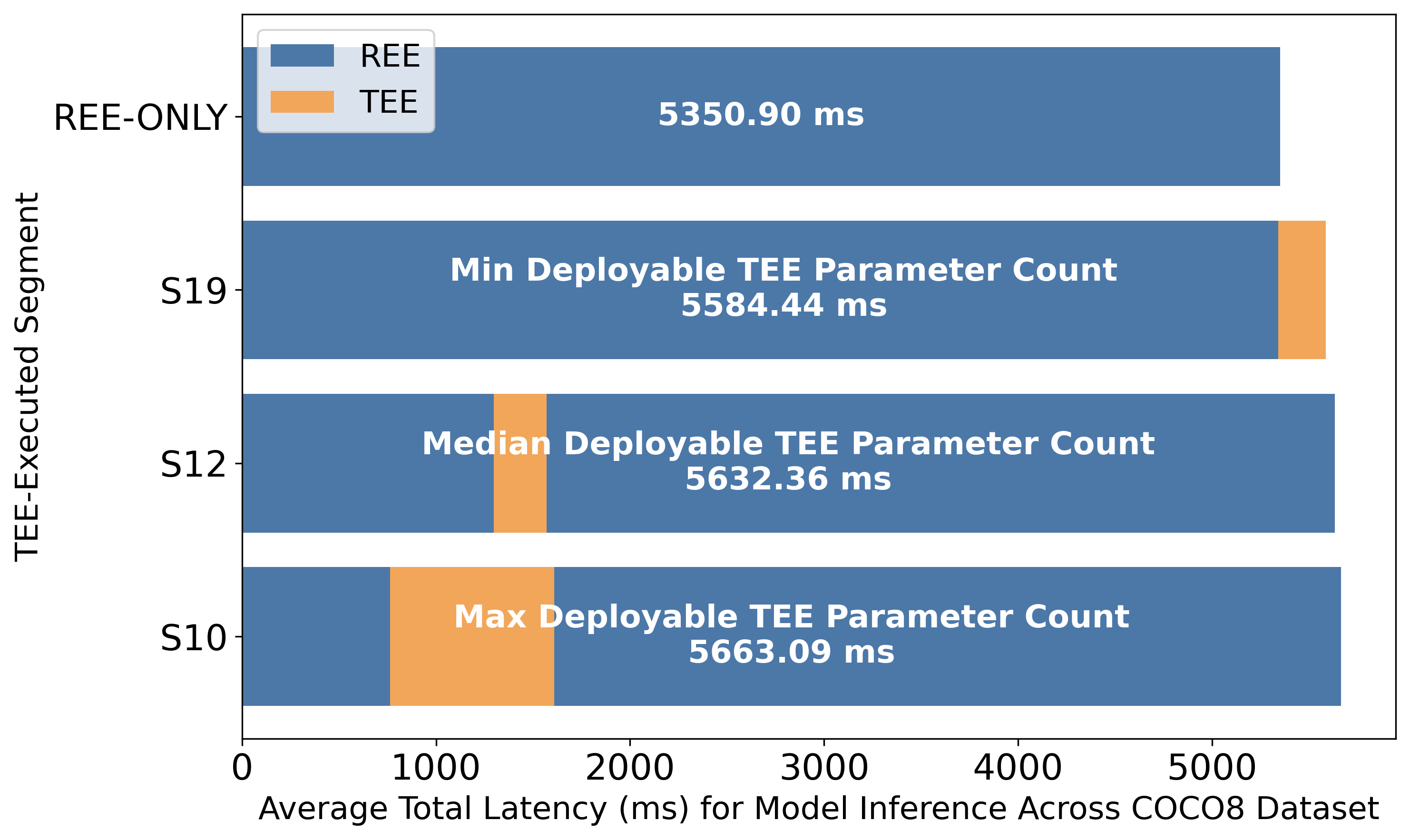}
        \caption{YOLO26n}
        \label{fig:YOLO26n_latency_representative_param_segs}
    \end{subfigure}

    \caption{\textbf{Inference Latency Breakdown for Representative TEE-Executed Parameter-Bearing Model Segments for (a) MNIST and (b) YOLO26n.}
    \normalfont Each bar shows the average total inference latency in milliseconds across the corresponding dataset, decomposed into Rich Execution Environment (REE) and Trusted Execution Environment (TEE) execution. TEE execution includes world-switching. REE-ONLY represents the baseline with no TEE execution. Values inside the bars indicate total average latency. For MNIST, segments 8, 10, and 7 correspond to the maximum (97.06\%), median (0.62\%), and minimum ($<$0.01\%) deployable TEE parameter counts, respectively. For YOLO26n, segments 10, 12, and 19 correspond to the maximum (1.51\%), median (0.17\%), and minimum ($<$0.01\%) deployable TEE parameter counts, respectively.
   These results illustrate the latency cost of TEE execution and show that cost depends on both the model and selected segment.
    }
    \label{fig:latency_representative_param_segs}
\end{figure}

Figures ~\ref{fig:MNIST_latency_representative_param_segs} and ~\ref{fig:YOLO26n_latency_representative_param_segs} show the average inference latency for REE-only execution and representative deployable partitions containing parameter-bearing segments. Latency is decomposed into REE and TEE execution, with the latter including the overhead of crossing between execution worlds.
For both models, REE-only execution provides the lowest measured latency, indicating that TEE-based execution introduces additional runtime overhead. This overhead could be attributed in part to the Secure Monitor calls, calling the construction/destruction of the trusted application in memory, and start/end of an inference session between the client application and trusted application for every inference. 
For MNIST, this overhead is particularly pronounced: even the parameter-lightest represented partition has an inference time 93\% greater than the REE baseline. However, the partition containing Segment 8, which holds 97.06\% of the model parameters, has only a small latency difference from the other represented TEE partitions. Thus, for the small MNIST model, the additional TEE execution overhead is substantial relative to the model's overall inference time.
YOLO26n shows a smaller relative difference between the REE-only execution and the representative TEE partitions. The parameter-light Segments 19 and 12 have similar latency to the largest deployable TEE partition containing Segment 10, despite its TEE execution occupying more of the inference latency. This suggests that latency depends not only on whether execution enters the TEE, but also on the computation performed within the selected partition.
Overall, these results show that TEE partitioning introduces a measurable latency cost, while the magnitude of that cost varies across partition configurations and models. Because the measured end-to-end latency combines TEE computation with REE/TEE transition overhead, the current measurements cannot independently attribute the additional latency to world-switching or TEE computation. Separating these costs is thereby important for future work when evaluating the latency-security tradeoff of individual partitions.




\section{Conclusion}
Our evaluation demonstrates that effective TEE-based model weight protection is a partition-selection problem constrained by model architecture, parameter distribution, TEE memory capacity, and inference latency. 
\textbf{\texttt{Onnx2TEE}} partitions models according to graph dependencies. However, the resulting segments can contain highly unequal number of parameters, as seen with 97.06\% of MNIST parameters and 96.96\% of YOLO26n parameters being concentrated in a single segment. Therefore, the number of segments placed in the TEE is not necessarily representative of the amount of model parameters protected. Parameter-dense segments may provide greater protection, yet consideration for their memory requirements must be made for partition deployment feasibility. Furthermore, results indicate that TEE memory depends not only on model weights and activations, but also on communicated boundary tensors, surprisingly meaning that adding a segment can at times reduce the overall memory requirement. These observations suggest that effective partition selection should jointly consider model weight coverage and resource requirements.

\textbf{\texttt{Onnx2TEE}} provides several practical advantages for split inference utilizing the TEE. Model-specific code generation avoids embedding a general-purpose runtime in the TEE, reducing the trusted code base. The framework also embeds model weights into the generated TEE application rather than transferring model weights from the CA to the TA during model inference. Use of the GlobalPlatfrom TEE API provides a path towards portability across compatible TEE implementations. 

\subsection{Limitations}
A few evaluation limitations remain. First, the current latency measurements combine TEE model inference computation with TA lifecycle and REE/TEE world-switching. This prevents an independent evaluation of TEE utilization for sub-model execution as opposed to loading and destroying the TA, responses to CA sessions, and other actions. This could be addressed by synchronizing a similar timing mechanism within the TEE's generated TA at different points of its life cycle using secure world logging. Because of its intended application toward edge devices, the software generated by our tool may likely run several inferences consecutively. Therefore, maintaining the TA in memory across inference experiments and separately measuring the switching overhead would provide a clearer characterization of the performance tradeoff. Additionally, the segmentation algorithm is dependent on model architecture and can produce segments too large for the available TEE memory. The current evaluation also only considers two models, limiting the generality of these observations.

Accounting for framework practicality, the model-specific code generation introduces additional developer consideration. Functions unavailable in the TEE environment may require custom implementations, and numerical differences between the REE and TEE implementations can affect inference. During development, for example, a Taylor-series approximation of \texttt{expf()} produced different YOLO26n behavior depending on where the approximation was applied, demonstrating that numerical consistency is important for generated TEE operators. 

\textbf{\texttt{Onnx2TEE}} inherits the security limitations of the TEE, including potential side-channel attacks and vulnerabilities in the trusted software stack. Moreover, because partition selection and model weights are incorporated into the generated binaries at compile time, changing the protected model region requires recompilation and redeployment.

\subsection{Future Work}
Future work should investigate automatic security-aware partition selection for proprietary models. While the current system exposes candidate partitions and evaluates their resource requirements, it does not distinguish between parameters based on their security value. Studying parameter importance, including changes introduced through domain-specific fine-tuning, could provide a basis for prioritizing model regions that offer the greatest protection when the complete model cannot fit within the TEE.  An approach based on results from such a study could automatically select partitions according to the security and resource requirements of a deployment rather than requiring a developer to manually choose the partition.

Future work should also investigate whether sensitive model regions can be distributed across multiple TEE-resident components rather than requiring a single contiguous protected partition. Such an approach could provide greater flexibility when security-sensitive parameters are separated by computation that is less sensitive or when no single contiguous region satisfies the available TEE memory constraints. This could motivate new segmentation strategies that jointly consider model dependencies, parameter sensitivity, and memory capacity when constructing multiple protected regions. 
More broadly, future work could examine whether protecting other proprietary model assets, such as model-specific post-processing, provides additional security benefits beyond protecting weights alone.

%
%
%
\bibliographystyle{IEEEtran}
\bibliography{references.bib}

@techreport{intel_sgx_programming_reference_2014,
  author       = {{Intel Corporation}},
  title        = {{Intel Software Guard Extensions Programming Reference}},
  institution  = {Intel Corporation},
  year         = {2014},
  number       = {329298-002},
  type         = {Programming Reference},
  url          = {https://www.intel.com/content/dam/develop/external/us/en/documents/329298-002-629101.pdf},
  note         = {Accessed: 2026-07-01}
}

@article{SGXexplained,
  title={Intel SGX explained},
  author={Costan, Victor and Devadas, Srinivas},
  journal={Cryptology ePrint Archive},
  year={2016}
}

@online{amd_sev,
  author  = {{Advanced Micro Devices, Inc.}},
  title   = {AMD Secure Encrypted Virtualization (SEV)},
  year    = {2026},
  url     = {https://www.amd.com/en/developer/sev.html},
  urldate = {2026-07-01},
  note    = {Developer documentation and specifications}
}

@techreport{arm_trustzone_2009,
  title        = {Arm Security Technology: Building a Secure System using TrustZone Technology},
  author       = {{Arm Limited}},
  institution  = {Arm Limited},
  year         = {2009},
  number       = {PRD29-GENC-009492},
  url          = {https://developer.arm.com/documentation/PRD29-GENC-009492/c},
  note         = {Accessed: 2026-07-01}
}

@inproceedings{costan2016sanctum,
  title={Sanctum: Minimal hardware extensions for strong software isolation},
  author={Costan, Victor and Lebedev, Ilia and Devadas, Srinivas},
  booktitle={25th USENIX Security Symposium (USENIX Security 16)},
  pages={857--874},
  year={2016}
}

@article{jauernig_trusted_2020,
	title = {Trusted {Execution} {Environments}: {Properties}, {Applications}, and {Challenges}},
	volume = {18},
	issn = {1558-4046},
	shorttitle = {Trusted {Execution} {Environments}},
	url = {https://ieeexplore.ieee.org/document/9041685},
	doi = {10.1109/MSEC.2019.2947124},
	number = {2},
	urldate = {2026-07-29},
	journal = {IEEE Security \& Privacy},
	author = {Jauernig, Patrick and Sadeghi, Ahmad-Reza and Stapf, Emmanuel},
	month = mar,
	year = {2020},
	pages = {56--60},
}

@misc{globalplatform2018tee,
  author       = {{GlobalPlatform}},
  title        = {Introduction to Trusted Execution Environments},
  year         = {2018},
  month        = may,
  howpublished = {\url{https://globalplatform.org/wp-content/uploads/2018/05/Introduction-to-Trusted-Execution-Environment-15May2018.pdf}},
  note         = {White paper. Accessed: 2026-07-29}
}

@misc{li_survey_2023,
	title = {A {Survey} of {Secure} {Computation} {Using} {Trusted} {Execution} {Environments}},
	url = {http://arxiv.org/abs/2302.12150},
	doi = {10.48550/arXiv.2302.12150},
	urldate = {2026-07-29},
	publisher = {arXiv},
	author = {Li, Xiaoguo and Zhao, Bowen and Yang, Guomin and Xiang, Tao and Weng, Jian and Deng, Robert H.},
	month = feb,
	year = {2023},
	note = {arXiv:2302.12150 [cs.CR]},
}

@article{munoz_survey_2023,
	title = {A survey on the (in)security of trusted execution environments},
	volume = {129},
	issn = {0167-4048},
	url = {https://doi.org/10.1016/j.cose.2023.103180},
	doi = {10.1016/j.cose.2023.103180},
	number = {C},
	urldate = {2026-07-30},
	journal = {Comput. Secur.},
	author = {Muñoz, Antonio and Ríos, Ruben and Román, Rodrigo and López, Javier},
	month = jun,
	year = {2023},
}

@article{tramer_stealing_nodate,
	title = {Stealing {Machine} {Learning} {Models} via {Prediction} {APIs}},
	language = {en},
	author = {Tramèr, Florian},
}

@inproceedings{cerdeira_sok_2020,
	address = {San Francisco, CA, USA},
	title = {{SoK}: {Understanding} the {Prevailing} {Security} {Vulnerabilities} in {TrustZone}-assisted {TEE} {Systems}},
	copyright = {https://ieeexplore.ieee.org/Xplorehelp/downloads/license-information/IEEE.html},
	isbn = {978-1-7281-3497-0},
	shorttitle = {{SoK}},
	url = {https://ieeexplore.ieee.org/document/9152801/},
	doi = {10.1109/SP40000.2020.00061},
	language = {en},
	urldate = {2026-08-03},
	booktitle = {2020 {IEEE} {Symposium} on {Security} and {Privacy} ({SP})},
	publisher = {IEEE},
	author = {Cerdeira, David and Santos, Nuno and Fonseca, Pedro and Pinto, Sandro},
	month = may,
	year = {2020},
	pages = {1416--1432},
}

@inproceedings{mo_darknetz_2020,
	title = {{DarkneTZ}: {Towards} {Model} {Privacy} at the {Edge} using {Trusted} {Execution} {Environments}},
	shorttitle = {{DarkneTZ}},
	url = {http://arxiv.org/abs/2004.05703},
	doi = {10.1145/3386901.3388946},
	urldate = {2026-08-03},
	booktitle = {Proceedings of the 18th {International} {Conference} on {Mobile} {Systems}, {Applications}, and {Services}},
	author = {Mo, Fan and Shamsabadi, Ali Shahin and Katevas, Kleomenis and Demetriou, Soteris and Leontiadis, Ilias and Cavallaro, Andrea and Haddadi, Hamed},
	month = jun,
	year = {2020},
	note = {arXiv:2004.05703 [cs.LG]},
	pages = {161--174},
}

@inproceedings{gangal_hybridtee_2020,
	title = {{HybridTEE}: {Secure} {Mobile} {DNN} {Execution} {Using} {Hybrid} {Trusted} {Execution} {Environment}},
	shorttitle = {{HybridTEE}},
	url = {https://ieeexplore.ieee.org/document/9358260/},
	doi = {10.1109/AsianHOST51057.2020.9358260},
	urldate = {2026-08-03},
	booktitle = {2020 {Asian} {Hardware} {Oriented} {Security} and {Trust} {Symposium} ({AsianHOST})},
	author = {Gangal, Akshay and Ye, Mengmei and Wei, Sheng},
	month = dec,
	year = {2020},
	pages = {1--6},
}

@article{hu_efftee_2025,
	title = {{EffTEE}: {Efficient} {Image} {Classification} and {Object} {Detection} on {Mobile} {Devices} {Using} {Trusted} {Execution} {Environments}},
	volume = {13},
	issn = {2169-3536},
	shorttitle = {{EffTEE}},
	url = {https://ieeexplore.ieee.org/document/10870347/},
	doi = {10.1109/ACCESS.2025.3538712},
	urldate = {2026-08-03},
	journal = {IEEE Access},
	author = {Hu, Bin and You, Junyong and Huang, Kuan and Xu, Meng and Liu, Dan and Ma, Sugang},
	year = {2025},
	pages = {31423--31441},
}

@article{steiakakis_trusted_2026,
	title = {Trusted {Yet} {Flexible}: {High}-{Level} {Runtimes} for {Secure} {ML} {Inference} in {TEEs}},
	volume = {6},
	copyright = {http://creativecommons.org/licenses/by/3.0/},
	issn = {2624-800X},
	shorttitle = {Trusted {Yet} {Flexible}},
	url = {https://www.mdpi.com/2624-800X/6/1/23},
	doi = {10.3390/jcp6010023},
	language = {en},
	number = {1},
	urldate = {2026-08-04},
	journal = {Journal of Cybersecurity and Privacy},
	publisher = {Multidisciplinary Digital Publishing Institute},
	author = {Steiakakis, Nikolaos-Achilleas and Vasiliadis, Giorgos},
	month = feb,
	year = {2026},
	pages = {23},
}

@misc{jocher2026ultralyticsyolo26unifiedrealtime,
  title = {Ultralytics YOLO26: Unified Real-Time End-to-End Vision Models},
  author = {Glenn Jocher and Jing Qiu and Mengyu Liu and Shuai Lyu and Fatih Cagatay Akyon and Muhammet Esat Kalfaoglu},
  year = {2026},
  eprint = {2606.03748},
  archivePrefix = {arXiv},
  primaryClass = {cs.CV},
  doi = {10.48550/arXiv.2606.03748},
  url = {https://arxiv.org/abs/2606.03748},
}

@misc{emmtrixemx-onnx-cgen_2026,
	title = {emmtrix/emx-onnx-cgen},
	copyright = {Apache-2.0},
	url = {https://github.com/emmtrix/emx-onnx-cgen},
	urldate = {2026-08-11},
	publisher = {emmtrix Technologies GmbH},
	month = jul,
	year = {2026},
	note = {original-date: 2026-01-09T16:00:20Z},
}

@misc{onnx,
  title        = {{ONNX}: Open Neural Network Exchange},
  author       = {{ONNX Contributors}},
  howpublished = {\url{https://github.com/onnx/onnx}},
  year         = {2026},
  note         = {Open standard for machine learning interoperability}
}

@inproceedings{abadi2016tensorflow,
  title={$\{$TensorFlow$\}$: a system for $\{$Large-Scale$\}$ machine learning},
  author={Abadi, Mart{\'\i}n and Barham, Paul and Chen, Jianmin and Chen, Zhifeng and Davis, Andy and Dean, Jeffrey and Devin, Matthieu and Ghemawat, Sanjay and Irving, Geoffrey and Isard, Michael and others},
  booktitle={12th USENIX symposium on operating systems design and implementation (OSDI 16)},
  pages={265--283},
  year={2016}
}

@inproceedings{he2016deep,
  title={Deep residual learning for image recognition},
  author={He, Kaiming and Zhang, Xiangyu and Ren, Shaoqing and Sun, Jian},
  booktitle={Proceedings of the IEEE conference on computer vision and pattern recognition},
  pages={770--778},
  year={2016}
}

@inproceedings{szegedy2015going,
  title={Going deeper with convolutions},
  author={Szegedy, Christian and Liu, Wei and Jia, Yangqing and Sermanet, Pierre and Reed, Scott and Anguelov, Dragomir and Erhan, Dumitru and Vanhoucke, Vincent and Rabinovich, Andrew},
  booktitle={Proceedings of the IEEE conference on computer vision and pattern recognition},
  pages={1--9},
  year={2015}
}

@INPROCEEDINGS{EdgeFlow,
  author={Hu, Chenghao and Li, Baochun},
  booktitle={IEEE INFOCOM 2022 - IEEE Conference on Computer Communications}, 
  title={Distributed Inference with Deep Learning Models across Heterogeneous Edge Devices}, 
  year={2022},
  volume={},
  number={},
  pages={330-339},
  doi={10.1109/INFOCOM48880.2022.9796896}}

@article{qararyah2021computational,
  title={A computational-graph partitioning method for training memory-constrained DNNs},
  author={Qararyah, Fareed and Wahib, Mohamed and Dikbay{\i}r, Do{\u{g}}a and Belviranli, Mehmet Esat and Unat, Didem},
  journal={Parallel computing},
  volume={104},
  pages={102792},
  year={2021},
  publisher={Elsevier}
}

@article{tarnawski2020efficient,
  title={Efficient algorithms for device placement of dnn graph operators},
  author={Tarnawski, Jakub and Phanishayee, Amar and Devanur, Nikhil and Mahajan, Divya and Nina Paravecino, Fanny},
  journal={Advances in Neural Information Processing Systems},
  volume={33},
  pages={15451--15463},
  year={2020}
}

\end{document}